\documentclass[aps,prl,twocolumn,groupedaddress,showpacs]{revtex4}

\usepackage{graphicx}

\begin{document}

\title{Experimental verification of fidelity decay:\\ From perturbative to Fermi 
Golden Rule regime}

\author{R. Sch\"afer}
\email[]{rudi.schaefer@physik.uni-marburg.de}
\affiliation{Fachbereich Physik, Philipps-Universit\"at Marburg, Renthof 5,
D-35032 Marburg, Germany}

\author{T. Gorin}
\affiliation{Max-Planck-Institut f\" ur Physik komplexer Systeme,
               N\" othnitzer Str.~38, 
               D-01187 Dresden, Germany}

\author{T. H. Seligman}
\affiliation{Centro de Ciencias F\'{\i}sicas, Universidad Nacional 
Aut\'{o}noma de M\'{e}xico, Campus Morelos, C.~P. 62251, 
Cuernavaca, Morelos, M\'{e}xico}

\author{H.-J. St\"ockmann}
\affiliation{Fachbereich Physik, Philipps-Universit\"at Marburg, Renthof 5,
D-35032 Marburg, Germany}

\date{\today}

\begin{abstract}
The scattering matrix was measured for a flat microwave cavity with 
classically chaotic dynamics. The system can be perturbed by small changes of 
the geometry. We define the ``scattering fidelity'' in terms of parametric 
correlation functions of scattering matrix elements. In chaotic systems and 
for weak coupling the scattering fidelity approaches the fidelity of the 
closed system. Without free parameters the experimental results agree with 
random matrix theory in a wide range of perturbation strengths, reaching from 
the perturbative to the Fermi golden rule regime.
\end{abstract}



\maketitle

\newcommand {\xb} {{\bf x}}
\newcommand {\xbd} {{\bf x^\dag}}
\newcommand {\yb} {{\bf y}}
\newcommand {\ybd} {{\bf y^\dag}}
\newcommand {\zn} {{\bf z}_n}
\newcommand {\znd} {{\bf z}_n^{\bf\dag}}
\newcommand {\ree} {R_\phi\left(E_1,E_2\right)}

\newcommand {\bet}{B}
\newcommand {\gam}{C}

\newcommand{\rmd}{{\rm d}}
\newcommand{\rmi}{{\rm i}}
\newcommand{\rme}{{\rm e}}
\newcommand{\eff}{{\rm eff}}
\newcommand{\inter}{{\rm int}}
\newcommand{\tot}{{\rm tot}}

\newcommand{\la}{\langle}
\newcommand{\ra}{\rangle}
\newcommand{\lla}{\left\langle}
\newcommand{\rra}{\right\rangle}

The stability of quantum motion has been a topic of increasing interest 
in recent years. In Ref.~\cite{Per84}, Peres proposed to consider the time 
evolution of wave packets governed by two slightly different Hamiltonians.
Starting from the same initial state, their overlap provides a natural 
measure for the stability of the quantum evolution. 
As ``fidelity'' and 
``quantum Loschmidt echo'', this quantity has since been investigated 
extensively (see Ref.~\cite{pro03b} and references therein).
Nowadays, it has become a standard benchmark 
for the reliability of quantum information processing~\cite{Nie00}. 
Following Ref.~\cite{pro03b}, one may define 
fidelity as $F(t)=|f(t)|^2$ and fidelity amplitude as
\begin{equation}\label{eq:fid}
f(t) = \la \psi(0) \left| U^\dagger(t) \, U^{\prime}(t) \right| \psi(0)\ra
\; ,
\end{equation}
where the unitary operators $U^{\prime}(t)$ and $U(t)$ describe the perturbed 
and unperturbed time evolution of an arbitrary initial state $\psi(0)$. 
Depending on the strength of the perturbation one can discern three regimes. 
In the perturbative regime, where time-independent perturbation theory
can be applied, the decay of the fidelity is Gaussian. For larger perturbations
a cross-over to exponential decay is observed, with a decay constant obtained
from Fermi's golden rule~\cite{Jac01b,Cer02}. For very strong perturbations 
the decay constant saturates at the classical Lyapunov exponent~\cite{Jal01}.

Since the first spin-echo experiment by Hahn~\cite{Hah50}, echo experiments
have been performed with many different quantum and classical wave systems 
(e.g.\ Ref.~\cite{Kur64,Buc00}). However, wave functions are usually not 
accessible to experiments, and only some reduced information is available, 
such as the nuclear induction averaged over the probe in a magnetic resonance 
experiment~\cite{Zha92,Pas95}, or the transmission between two antennas in a 
microwave or ultrasound experiment~\cite{Ler04,Der95,Lob03a}.

Here, we report on the experimental measurement of fidelity decay in a
flat electromagnetic cavity, using the equivalence of 
Helmholtz and stationary Schr\"odinger equation~\cite{Stoe99}. 
Instead of following the evolution of wave packets, we measure stationary 
spectra of scattering matrix elements, separately, for the perturbed and the 
unperturbed system. Then, for a given scattering matrix element, we compute 
the Fourier transform of the cross-correlation function between the two 
spectra. After an appropriate normalization, this defines the {\em scattering 
fidelity} amplitude. 
%
Averaging this quantity over a large number of uniformly distributed
antennas with small transmission yields the standard fidelity
amplitude. Yet for integrable systems, this may still lead to system specific results.
For chaotic systems, by contrast, all antenna positions become equivalent
and measurements for a few of them are sufficient. This statement is made
more precise in~\cite{schae}. Its validity is supported by our
experimental results which agree with the universal 
prediction of random matrix theory~\cite{pro03b,gor04}.
%
%

%
%
Our microwave experiments are adequately described by the statistical
scattering theory~\cite{Stoe99,Mah69}, if absorption is taken into account~\cite{Sch03a}. 
The scattering matrix for a billiard with two antennas 
can be written as:
\begin{equation}
S_{ab} (E) = \delta_{ab} 
   - \rmi\, {V^{(a)}}^\dagger\; \frac{1}{E-H_{\rm eff}}\; V^{(b)}  \; .
\label{eq:scatt}
\end{equation}
Here $H_{\rm eff}= H_\inter - (\rmi/2)\, V\, V^\dagger$ is the effective 
Hamiltonian of the open system and $H_\inter$ is the Hamiltonian of the closed 
billiard.  
The column vectors of $V$, denoted by $V^{(a)}$, contain the information on the coupling to the antennas at positions $\vec r_a$. For antenna diameters small 
compared to the wavelength, the coefficients $V_{ja}$ 
are proportional to $\psi_j(\vec r_a)$, the wave functions of the closed system 
at the antenna positions.

\begin{figure}
\begin{center}
\includegraphics[width=0.48\textwidth]{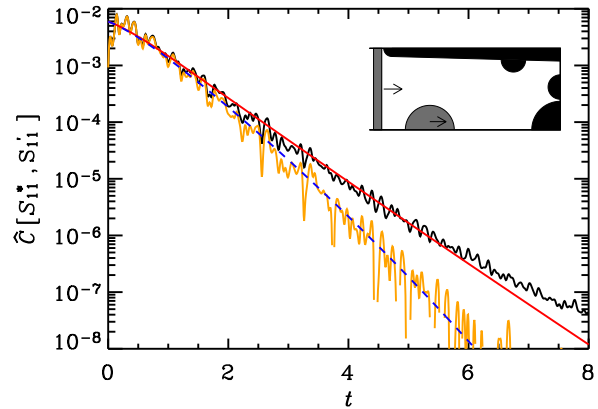}  \\[-2ex] 
 \caption{(Color online): Logarithmic plot of the correlation function 
$\hat C[S_{11}^\ast,S_{11}^\prime]$ for $\nu=5$--6~GHz, $l=1$mm and 
$\lambda=0.047$. The experimental results for the auto correlation are shown 
in black, while the correlation of perturbed and unperturbed system are 
shown in gray / orange. The smooth solid curve corresponds to the 
theoretical auto-correlation function, and the dashed curve to the product 
of auto-correlation function and fidelity amplitude. 
The insert shows the billiard geometry used. Moveable parts are marked with an
arrow. See text for dimensions.}
\label{fig:auto}
\end{center}
\end{figure}

Consider the cross-correlation function between an S-matrix element of the 
perturbed and the unperturbed system in the time domain:
\begin{equation}
\hat C[S_{ab}^\ast,S_{ab}^\prime](t)  \propto \la \hat S^\ast_{ab}\, (t)\; 
\hat S^\prime_{ab}\, (-t) \ra \, ,
\label{eq:echocor}
\end{equation}
where $S'_{ab}(E)$ is given by Eq.(\ref{eq:scatt}), but with $H_\inter$ 
replaced by $H'_\inter$. The brackets denote an energy window and/or ensemble 
average. Note that $\hat C[S^\ast_{ab},S_{ab}^\prime](t)$ 
describes a kind of echo-dynamics, which is similar to the quantum echo 
defined in Eq.~(\ref{eq:fid}), but
decays even without any perturbation.
We therefore use the autocorrelation function for a heuristic 
normalization and define the scattering fidelity amplitude as
\begin{equation}
f_{ab}(t)=\frac{\hat C[S_{ab}^\ast,S_{ab}^\prime](t) }
  {\sqrt{\hat C[S_{ab}^\ast,S_{ab}](t)\;
   \hat C[S_{ab}^{\prime\ast},S_{ab}^\prime](t) }} \; .
\label{eq:def_fscat}
\end{equation}
Reflection and transmission measurements have been performed in a flat 
microwave cavity, with top and bottom plate parallel to each 
other~\cite{Kuh00b}. The cavity is quasi-two-dimensional for frequencies
$\nu < c/(2h)$, where $h$ is the height of the billiard.
The billiard is shown in the insert of Fig.~\ref{fig:auto}. It 
consists of a rectangular cavity of length $L= 438 $\,mm, width 
$B=200$\,mm and height $h=8$\,mm, a quarter-circle insert of radius 
$R_1=70$\,mm, and a half-circle insert of radius $R_2=60$\,mm placed on the 
lower side. The position of the latter was changed in steps of 20\,mm to 
generate an ensemble average over 15 different systems.
Additional elements were inserted into the billiard to suppress bouncing-ball 
resonances: two half-circle inserts with radius $R_3=30$\,mm, and a slope on 
the upper boundary.  The perturbation of the system was achieved by varying 
the length $L$ in steps of $l=n\cdot 0.2$\,mm, with $n=1$--$10$.
The change of area and surface due to the 
shift of the billiard wall, was taken into account by unfolding the 
spectra to a mean level distance of one. The frequency window of the Fourier 
transforms was 1\,GHz wide, and a Welch filter was applied. In this range the 
antenna coupling and the wall absorption are approximately constant. 

We compare the experimental results with the random matrix prediction
\begin{equation}\label{eq:fd_lin}
  f(t)= \exp\!\!\left[- 4 \pi^2 \lambda^2 \left(\!
     t^2\! +\frac{t}{2}- \!\!\int_0^t\!\! \int_0^\tau\!\!\!
b_2(\tau^\prime)\; \rmd \tau^\prime\, \rmd \tau\!\! \right)\!\!\right]\, .
\end{equation}
This expression is obtained by 
exponentiation of the linear response result, 
thus incorporating the known behavior in both, the perturbative and the Fermi golden rule regime~\cite{gor04}.
Here, $\lambda$ is the perturbation strength, and $b_2(t)$ is the two-point 
form factor for the Gaussian orthogonal ensemble.
We use dimensionless units, where the Heisenberg time $t_H=\hbar/\Delta$ is equal to one and $\Delta$ denotes the mean level spacing.

%
%

Figure~\ref{fig:auto} shows on a logarithmic scale the cross-correlation 
function~$\hat C[S_{11}^\ast,S_{11}^\prime](t)$ given in 
Eq.~(\ref{eq:echocor}) together with the  autocorrelation 
function~$\hat C[S_{11}^{\ast},S_{11}](t)$. The latter
agrees with the corresponding theoretical autocorrelation function, 
calculated in~\cite{Sch03a}. The parameters for the wall absorption and the 
coupling of the antennas have been obtained according to the same reference.
With increasing time, $\hat C[S_{11}^\ast,S_{11}^\prime](t)$ deviates more 
and more from the autocorrelation function. 
This deviation contains the essential information on echo dynamics.

\begin{figure}
\begin{center}
\includegraphics[width=0.48\textwidth]{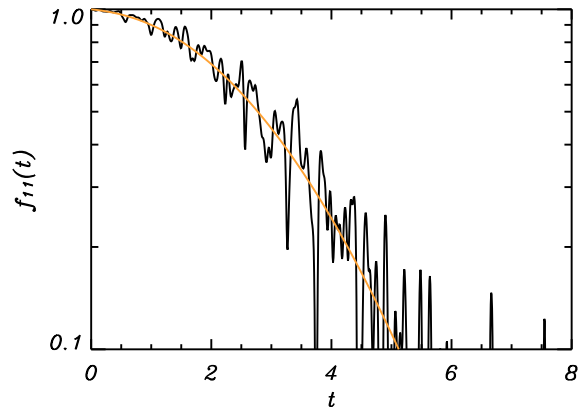}   \\[-2ex] 
 \caption{(Color online): Logarithmic plot of the scattering fidelity 
amplitude $f_{11}(t)$  for $\nu=5$--6~GHz, $l=1$mm and $\lambda=0.047$. The 
smooth curve shows the linear-response result for the  fidelity
amplitude $f(t)$, where the perturbation strength $\lambda$ was obtained 
from the variance of the level velocities.}
\label{fig:fdamp}
\end{center}
\end{figure}

For the frequency range shown in Fig.~\ref{fig:auto}, the perturbation 
strength $\lambda$ was determined directly from the measured spectra via the 
variance of the level velocities.
The dashed curve in Fig.~\ref{fig:auto} is a product of the theoretical 
autocorrelation function and the fidelity amplitude (Eq.~(\ref{eq:fd_lin})) 
of the closed system.
The experimental result for the cross-correlation function 
(Eq.~(\ref{eq:echocor})) agrees perfectly over six orders of magnitude with 
the linear-response expression without any free parameter.
This justifies our definition of the scattering fidelity amplitude $f_{ab}(t)$ in 
Eq.~(\ref{eq:def_fscat}).

Figure~\ref{fig:fdamp} shows $f_{ab}(t)$, computed from the
experimental cross- and autocorrelation functions according to 
Eq.~(\ref{eq:def_fscat}). There are two advantages for using the experimental 
and not the theoretical autocorrelation function: The computed quantity does 
not depend on theoretical assumptions, and the influence of non-generic 
features, visible in the correlation functions, is reduced.

\begin{figure}
\begin{center}
 \includegraphics[width=0.45\textwidth]{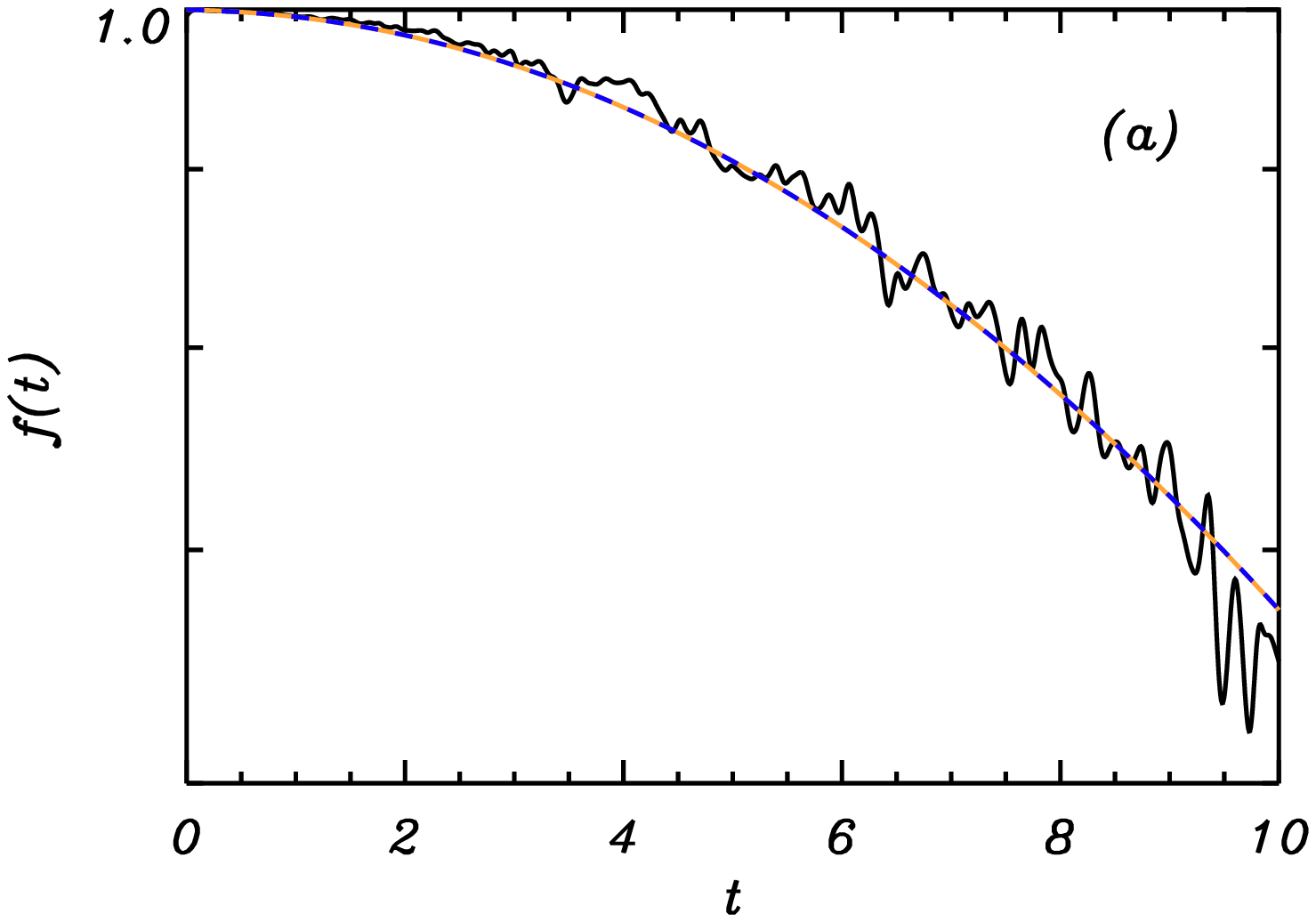} \\[-4ex] 
 \includegraphics[width=0.45\textwidth]{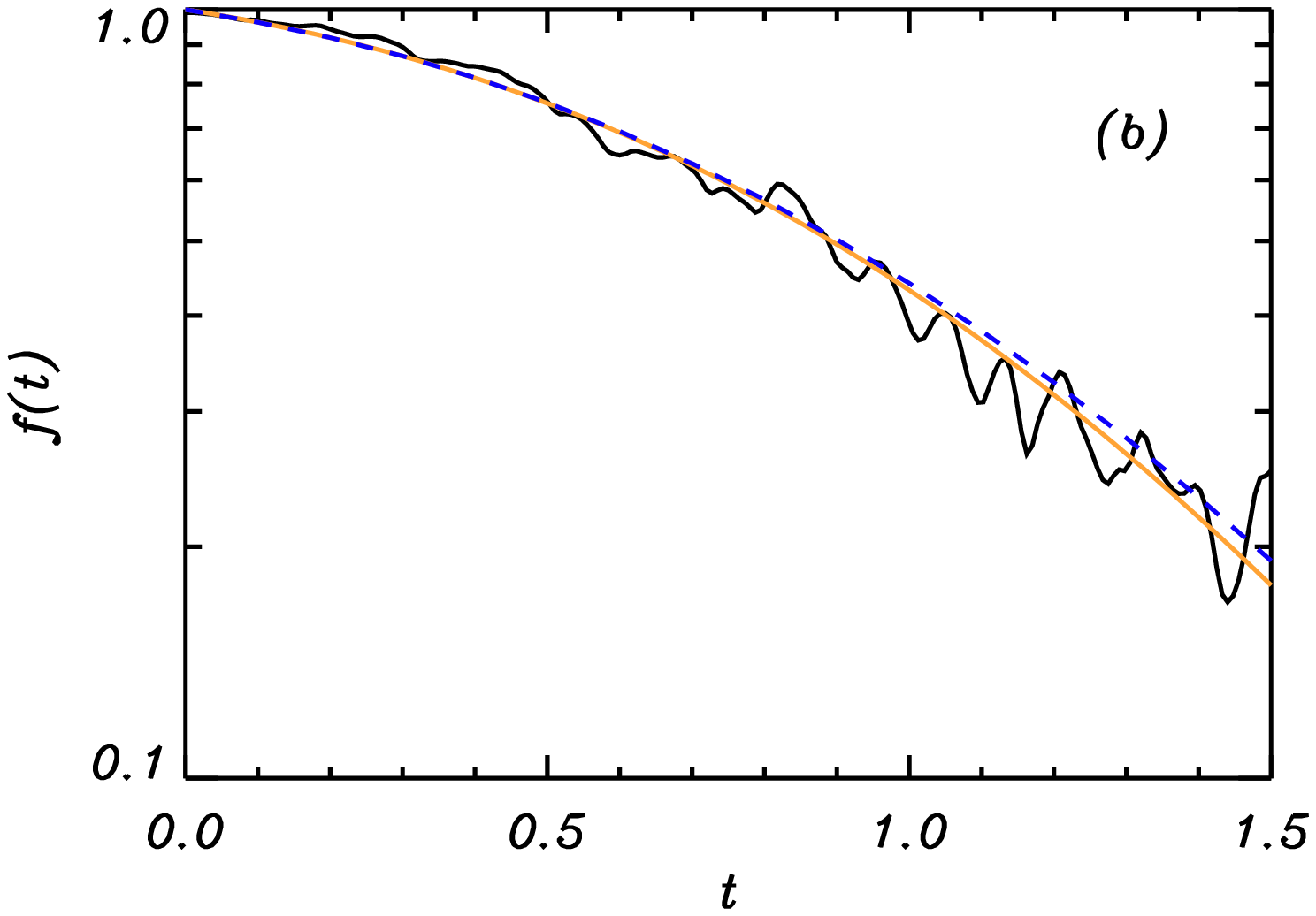} \\[-4ex] 
 \includegraphics[width=0.45\textwidth]{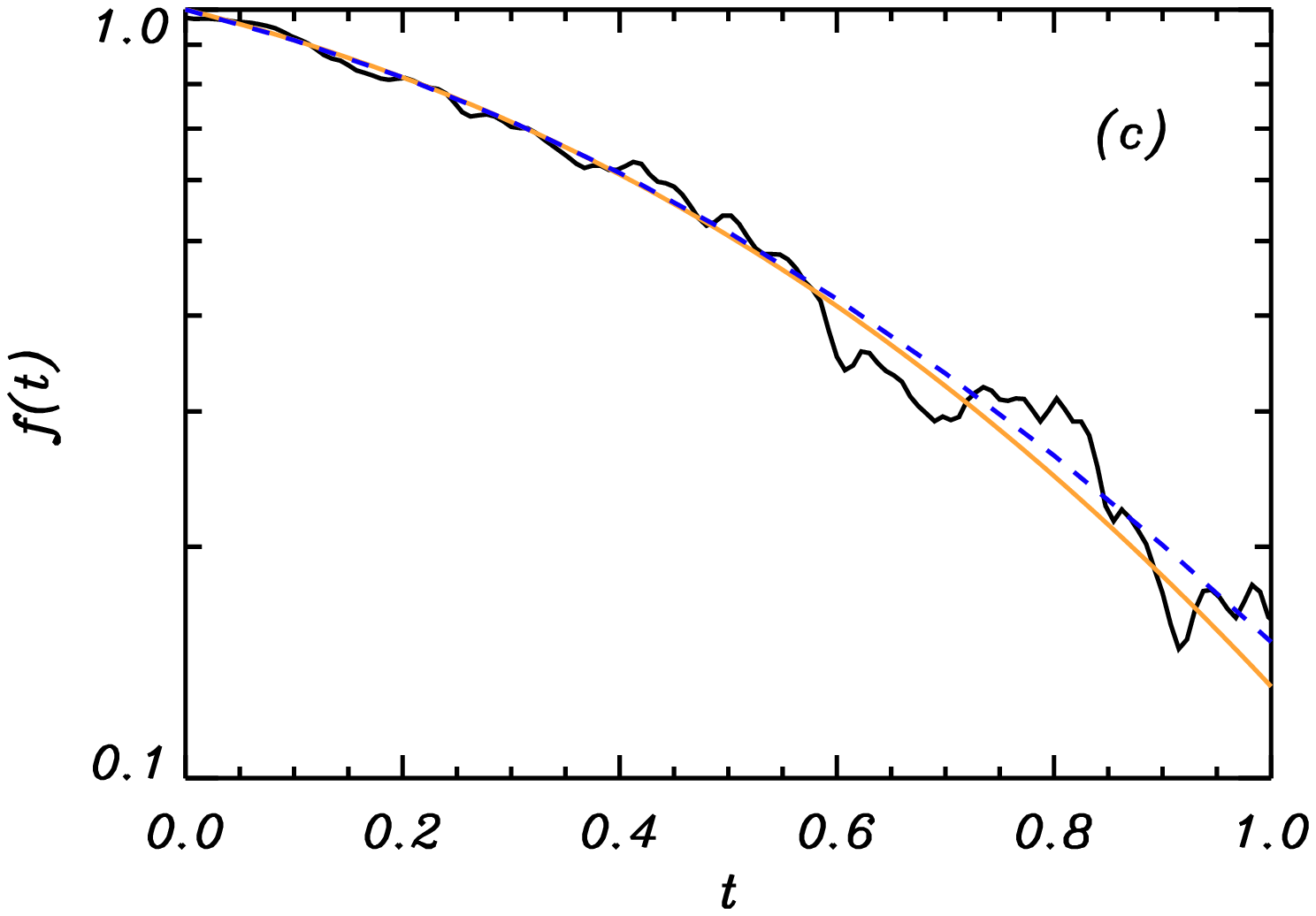}  \\[-3ex] 
\caption{Logarithmic plot of the fidelity amplitude $f(t)$ for three different 
perturbation strengths. 
Linear response result (smooth solid gray line), exact theoretical result
(dark dashed line), and experimental result (solid dark line).
The perturbation parameter $\lambda$ has been fitted to each 
experimental curve. The parameters were $\nu=3$--4~GHz, $l=0.4$mm, 
$\lambda=0.01$ for (a),  $\nu=13$--14~GHz, $l=0.6$mm, $\lambda=0.13$ for (b), 
and  $\nu=16$--17~GHz, $l=0.8$mm, $\lambda=0.21$  for (c).}
\label{fig:fidelity}
\end{center}
\end{figure}

We now study the dependence of the fidelity decay on the perturbation strength.
In our experiment $\lambda$ varies from $\lambda=0.01$ for $n=1$ and 
$\nu=3$--$4$\,GHz up to $\lambda=0.5$ for $n=10$ and $\nu=17$--$18$\,GHz.
Figure~\ref{fig:fidelity} shows the scattering fidelity amplitude for three 
different frequency windows. Here, $\lambda$ has been 
fitted to the experimental curves, as its determination from the level 
dynamics is time consuming and for strong perturbations not always feasible.  
To improve statistics, experimental results for $f_{11}$, $f_{22}$ and $f_{12}$ have been superimposed.

For the random matrix model one expects a transition from linear to 
quadratic decay near the Heisenberg time. 
In the perturbative regime, the linear term in the exponential is still 
close to one and we observe 
Gaussian decay of the fidelity amplitude, as seen in 
Fig.~\ref{fig:fidelity}(a). 
With increasing perturbation strength the linear term becomes more pronounced, 
leading to the 
Fermi golden rule regime~\cite{Jac01b,Cer02}. 
The (exponentiated) 
linear-response formula (\ref{eq:fd_lin}) 
agrees very well with experiment throughout the range. 
Recently, an exact solution for the random matrix model 
proposed in \cite{gor04} 
has been obtained using supersymmetry techniques \cite{stoe04b}. 
This result is shown as dashed lines in Fig.~\ref{fig:fidelity}. For the 
accessible perturbation strengths the experiment does not allow to distinguish 
between the linear-response and the exact result.

In billiard systems, the parameter variation is not due to a change of the 
Hamiltonian, but of the boundary condition. It was shown in chapter~5 of 
Ref.~\cite{Stoe99} that both situations are equivalent. For the case of a 
parameter variation in the billiard due to a shift of a straight wall 
the matrix element of the equivalent perturbation reads
\begin{equation}
(H_1)_{nm}=l \int_0^B \left. \frac{\partial \psi_n(x,y)}{\partial x} 
   \frac{\partial \psi_m(x,y)}{\partial x}\right|_{x=0}\, dy \, ,
\end{equation}
where $l$ is the shift of the wall (in $x$-direction) and $B$ is the length 
of the shifted wall. The perturbation strength according to the random 
matrix model in~\cite{gor04} is given by the variance of the off-diagonal 
matrix elements:
\begin{equation}
\lambda^2 = \left< \left[ (H_1)_{nm} \right]^2 \right> \,.
\end{equation}
Using Berry's conjecture of the superposition of plane waves~\cite{Ber77a} we 
can derive for large wave numbers $k$:
\begin{equation} \label{eq:scaling}
\lambda^2 = \frac{2 L}{3 \pi^3} k^3 l^2 = \frac{16 L}{3 c^3} \nu^3 l^2 \, .
\end{equation}
In Ref.~\cite{Leb99}, the same expression has been obtained 
using periodic orbit theory and the ergodicity assumption.

\begin{figure}
\begin{center}
\includegraphics[width=0.45\textwidth]{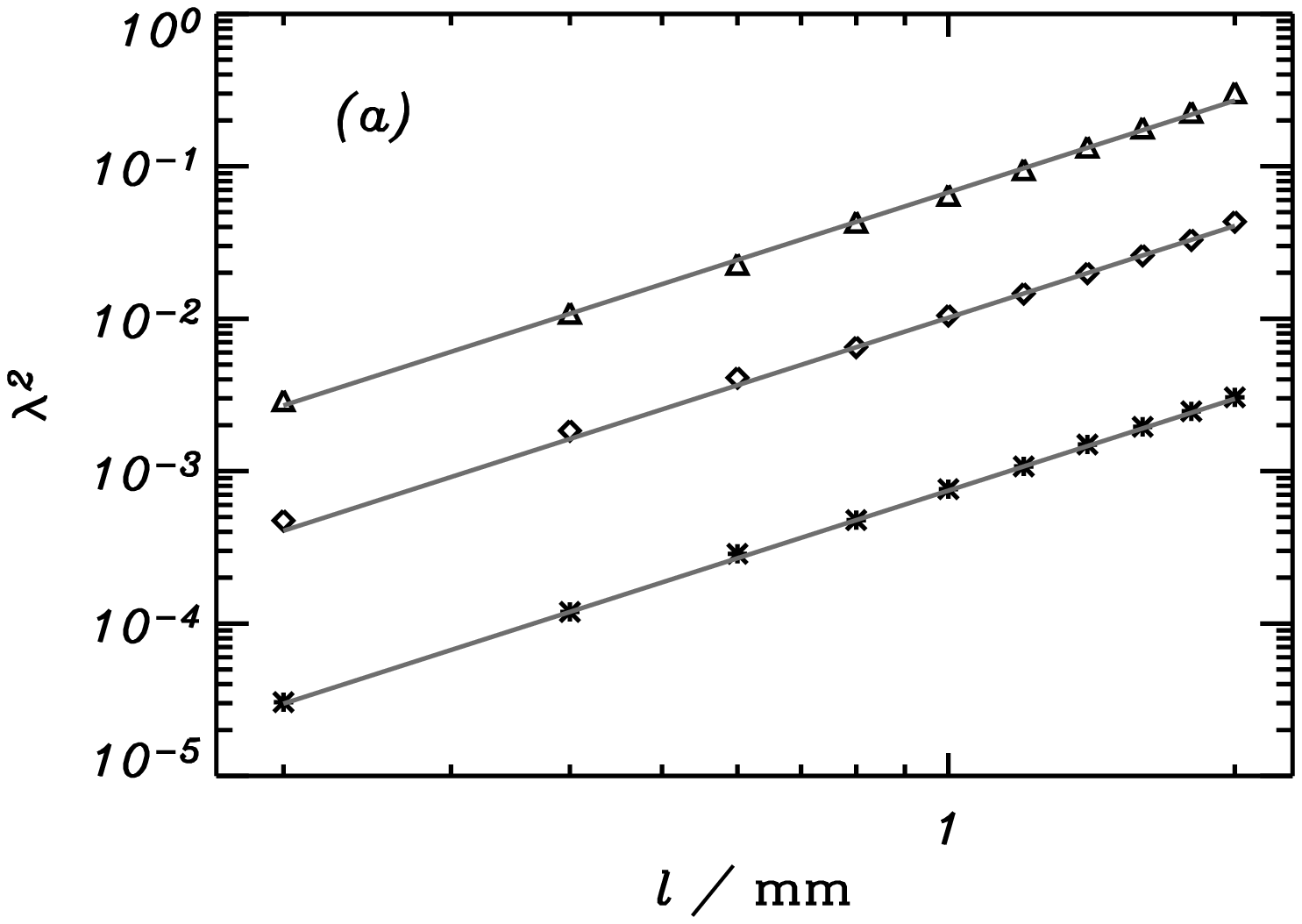} \\[-4ex] 
\includegraphics[width=0.45\textwidth]{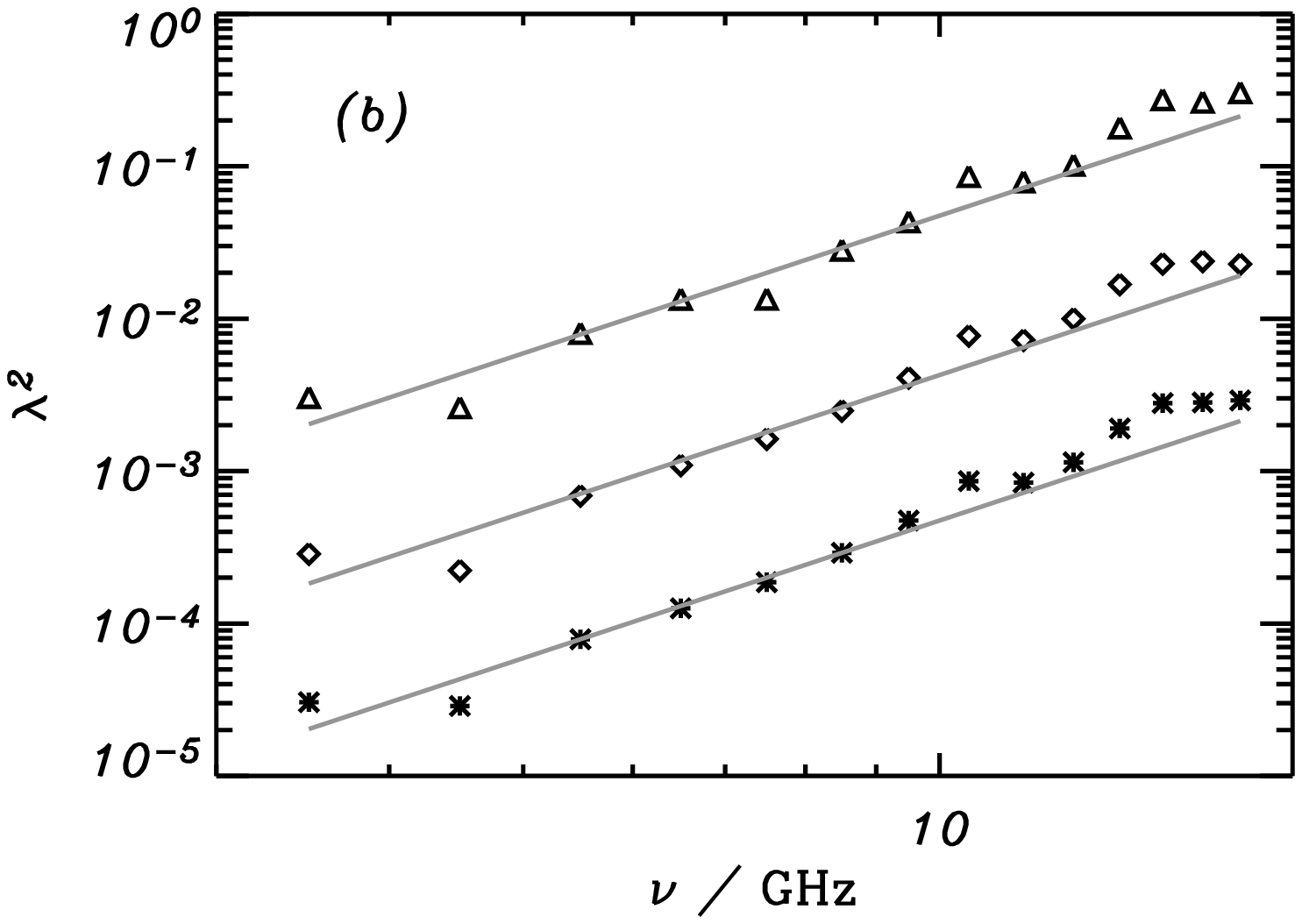} \\[-2ex] 
\caption{(a) Perturbation strength $\lambda^2$ as a function of the shift $l$ 
of the billiard wall for the frequency windows $\nu=3$--4 (stars), 9--10 
(diamonds) and 16--17\,GHz (triangles). The slope of the straight lines is 2.
(b) $\lambda^2$ as a function of the frequency $\nu$ for $l=0.2$ (stars), 0.6
(diamonds) and 2.0\,mm (triangles). The slope of the straight lines is 3.}
\label{fig:scale}
\end{center}
\end{figure}

Figure~\ref{fig:scale} shows the experimental perturbation strength
$\lambda^2$ as a function of the shift  $l$ of the billiard wall for three
different frequency regimes (a), and as a function of the frequency, for
three different shifts (b). We observe excellent agreement with the
scaling $\lambda^2 \propto l^2\, \nu^3$ predicted in 
Eq.~(\ref{eq:scaling}). 
This can be used to average all experimental data as a function of the
scaled variable $4\pi^2\lambda^2 C(t)$, thus reducing the fluctuations
almost completely~\cite{schae}.
In spite of the correct scaling, the experimental prefactor is about three
times smaller than predicted.
In cases where we determined the  variance of level velocities 
directly from the measured spectra, we found the same discrepancy. The 
deviation is caused by the fact that we are far from the semiclassical limit, 
for which Eq.~(\ref{eq:scaling}) was derived. 
Additional numerical studies for the Sinai billiard by H. Schanz~\cite{schanz} 
substantiate this explanation.

We have shown that the fidelity amplitude $f_{ab}(t)$ of scattering matrix 
elements $S_{ab}$ is an easily accessible quantity. In our experiment it 
approximates the ordinary fidelity amplitude $f(t)$. The relevant parameters 
are well controlled. This enabled us to verify the theoretical results for 
fidelity decay covering the range from the perturbative to the Fermi golden 
rule regime without any free parameter.
For strong perturbations the exact solutions of 
random matrix theory predict a revival of the fidelity amplitude at the 
Heisenberg time, whereas semiclassics predicts an exponential decay with 
the Lyapunov exponent up to the Ehrenfest time \cite{Jal01}. It remains an
open question, whether these features can be verified in a future microwave 
experiment.

\begin{acknowledgments}
T Prosen, U Kuhl and H Schanz are thanked for helpful discussions taking 
place in part on occasion of a workshop at the Centro Internacional de 
Ciencias in Cuernavaca, Mexico. T.H.S. acknowledges
support under the grants DGAPA 10803 and CONACyT 41000-F.
The experiments were supported by the Deutsche Forschungsgemeinschaft.
\end{acknowledgments}


\end{document}